# DETERMINISTIC OBLIVIOUS DISTRIBUTION (AND TIGHT COMPACTION) IN LINEAR TIME

ENOCH PESERICO, UNIV. PADOVA

ABSTRACT. In an array of $n$ elements, $m$ positions and $m$ elements are *marked*. We show how to permute the elements in the array so that all marked elements end in marked positions, in time $O(n)$ (in the standard word-RAM model), deterministically, and *obliviously* – i.e. with a sequence of memory accesses that depends only on $n$ and not on which elements or positions are marked.

As a corollary, we answer affirmatively to an open question about the existence of a deterministic oblivious algorithm with $O(n)$ running time for tight compaction (move the $m$ marked elements to the first $m$ positions of the array), a building block for several cryptographic constructions. Our $O(n)$ result improves the running-time upper bounds for deterministic tight compaction, for randomized tight compaction, and for the simpler problem of randomized loose compaction (move the $m$ marked elements to the first $O(m)$ positions) - until now respectively $O(n \lg n)$, $O(n \lg \lg n)$, and $O(n \lg^* n)$.

1. INTRODUCTION

In an array of $n$ words, each of $b \geq \lg n$ bits, $m$ positions and $m$ words are *marked*. The goal of *oblivious distribution* is to permute the words in the array so that all marked words end in marked positions, with a sequence of memory accesses that is data oblivious (i.e. depends only on $n$ and not on which words or positions are marked). Oblivious distribution and its special case of *oblivious tight compaction* (permute the array so as to move $m$ marked words to the first $m$ positions) are building blocks of several cryptographic constructions, e.g. [3, 9, 10, 11, 15, 18]. Oblivious tight compaction also automatically yields a solution for the simpler, but equally important [5, 10] problem of oblivious $\ell$-*loose compaction* (permute the array so as to move $m$ marked words to any $m$ of the first $m\ell$ positions). We refer the reader to [4, 10] for a far more comprehensive review of the applications of oblivious tight and loose compaction.

We show how to perform oblivious distribution deterministically and in time $O(n)$ in the standard word-RAM model [12], improving to $O(n)$ the best previous running times for oblivious deterministic tight compaction, randomized tight compaction, and randomized $O(1)$−loose compaction – respectively $O(n \lg n)$ [1], $O(n \lg \lg n)$ [4] [2], and $O(n \lg^* n)$ [10] [3].

Our work is organized as follows. Section 2 briefly reviews the (well-known) existence and constructibility of families of pseudorandom graphs, in which the number of edges between any two sets of vertices is roughly the same one would have in a random graph with the same edge density [7, 14]. These are, put simply, expanders; but the formulation in terms of pseudorandomness captures the property essential to us, making our proofs more intuitive and our constructions more modular (e.g. they become significantly easier to re-randomize). Section 3 exploits these graphs to carry out $O(1)$−loose compaction deterministically and in linear time. The key idea to achieve obliviousness without sacrificing linearity is to decouple the task of moving the data from that of computing their destinations, and carry out the latter at "multiple scales" – compressing the computation's own data when operating at "small scale", and amortizing the computation's cost over large data transfers when operating at "large scale". Section 4 shows how pseudorandom graphs can be used to translate, deterministically and with a constant multiplicative overhead, *any* oblivious $O(1)$−loose compaction scheme into a scheme for oblivious distribution, proving our main result. Section 5 concludes with a brief discussion about the practical implications of our work: in particular, it examines the two main sources of the large constants hidden in the asymptotic notation (and how they may be reduced), and looks at how our scheme would fare in more realistic memory models incorporating parallelism, non-uniform access cost, block transfer, and pipelining.

---

[1]This can be achieved through a straightforward simulation of a butterfly network [2]. Note that there exist $n$−node, $O(1)$-degree "superconcentrator" networks [20] that can effect tight compaction/distribution; but they still operate in $\Omega(\lg n)$ steps, and unlike the butterfly, they keep some nodes active for more than 1 step. Even in cases when the sum over all steps of active nodes is $O(n)$, *which* nodes are active depends on the inputs – so tight compaction/distribution via any natural simulation of these networks is either non-oblivious, or requires $\Omega(n \lg n)$ operations.

[2]More precisely, [4] guarantees failure probability inverse superpolynomial in $\lambda$ with running time $O(n \lg \lg \lambda)$, i.e. $O(n \lg \lg n)$ for $\lambda = n$. This compares favourably with the worse (and less flexible) inverse $poly(n)$ probability of failure when simulating in time $O(n \lg \lg n)$ the probabilistic sorting networks of [16] through a random intermediate permutation [21].

[3]With probability of failure inverse polynomial in $n$ – unlike [4] above.

2. A flavour of pseudorandomness in multigraphs.

The rest of the paper assumes we know, for a sufficiently small $\epsilon > 0$ and some bounded $d_\epsilon$, how to construct for *any* power-of-2 $N$ in time $O(N)$ a $d-$regular balanced bipartite multigraph of $2N$ vertices satisfying a simple pseudorandomness property: in a nutshell, the number of edges $e(U, V)$ between any two subsets $U, V$ of its partitions must be sufficiently to what one would expect from a random graph of the same edge density [7, 14]. More formally:

**Definition 1.** *A $d_\epsilon-$regular bipartite multigraph $G_N(L, R, E)$ with $|L| = |R| = N$ satisfies the **DISC**($\epsilon d_\epsilon$) property if for all $U \subseteq L, V \subseteq R$ we have:*

$$\left| e(U, V) - \frac{d_\epsilon}{N} |U||V| \right| \leq \epsilon d_\epsilon \sqrt{|U||V|} \tag{1}$$

For all arbitrarily small $\epsilon > 0$ there are sufficiently large $d_\epsilon$ for which these graphs are known to exist and to be constructible in linear time; the purpose of this section is to briefly review the literature showing *one way* in which they can be obtained. One may recognize in Equation 1 the formula of the Expander Mixing Lemma [1] for (balanced) bipartite graphs [7]; and indeed a route to obtain multigraphs adequate for our purposes is that of [20] via spectral expansion, that we follow below for its simplicity and familiarity to many readers. We do stress however that it is not the only possibility: ultimately, the rest of the paper uses Definition 1 in a black-box fashion, so any linear-time construction of multigraphs of every size that is a power of 2 satisfying the definition works – and there are indeed a number of more complex but more efficient (in terms of how small $d_\epsilon$ can be made) solutions that are readily available in the literature, or require only a modicum of adaptation.

**Step 0:** $(N, d, \lambda)$ **graphs.** Let us call for brevity a balanced bipartite $d-$regular multigraph of $N$ vertices per partition, whose $N \times N$ biadjacency matrix is irreducible with second largest eigenvalue modulus equal to $\lambda$, an $(N, d, \lambda)$ graph; and noting that the largest eigenvalue of such a graph is exactly $d$, let us refer to $d/\lambda$ as its (multiplicative) eigengap.

**Step 1: dense infinite families of $(N, d, \lambda)$ graphs [13].** [13] shows how to construct for every square $N$, of an $(N, 8, 5\sqrt{2} < 8)$ graph in time $O(N)$. The (irreducible) biadjacency matrix is written as $\mathbf{M}^T + \mathbf{M}$, where $\mathbf{M}$ and $\mathbf{M}^T$ are each the sum of 4 permutation matrices, one for every edge of each vertex; this guarantees 8-regularity. Also, the single non-zero index of each row/column of each permutation matrix can be determined with $O(1)$ elementary operations, so the entire edge set is constructible in linear time. Proving the eigengap requires a fairly sophisticated analysis beyond the scope of this brief review.

**Step 2: boosting the eigengap [20].** Taking the $k^{th}$ power of the biadjacency matrix of an $(N, d, \lambda)$ graph yields an $(N, d^k, \lambda^k)$ graph – i.e. the eigengap $(d/\lambda)^k$ then be made arbitrarily high by choosing a sufficiently high but still bounded $k$ (note that irreducibility of a matrix guarantees irreducibility of its powers). Construction time remains linear in $N$, since moving from the $(i-1)^{th}$ to the $i^{th}$ power of the biadjacency matrix requires $O(1)$ operations for each of the $d$ edges that replace an edge in the previous matrix – so the total time per vertex is $\sum_{i=1}^{k} O(d^i) = O(d^k) = O(1)$.

**Step 3: padding the family [20].** Given a family of $(N, d^k, \lambda^k)$ graphs for all power-of-4 $N$ we can obtain a family of $(N, 2d^k, 2\lambda^k)$ graphs for all power-of-2 $N$: in other words, we can "fill the gaps" in the family by doubling the degree while maintaining the eigengap unchanged. From the $(N, d^k, \lambda^k)$ graph $H_M$ we obtain an $(N, 2d^k, 2\lambda^k)$ graph $G_N$ simply by replacing every edge of $H_N$ with two edges between the same vertices; and we obtain a $(2N, 2d^k, 2\lambda^k)$ graph $G_{2N}$ by taking two copies of $H_N$, and adding an edge between $u$ in the first and $v$ in the second if there is already edge between $u$ and the homologue of $v$. It is easy to see that both $G_N$ and $G_{2N}$ are $2d^k-$regular, and that they can be constructed from $H_N$ in time $O(N)$ doubling all eigenvalues of the biadjacency matrix. It is also easy to see that the biadjacency matrix remains irreducible (in the case of $G_{2N}$ it is formed by 4 identical blocks, each an irreducible matrix).

**Step 4: the Expander Mixing Lemma (for bipartite multigraphs)** [1, 8]. Finally, one can apply to any $N, 2d^k, 2\lambda^k$ graph the Expander Mixing Lemma [1, 8] tailored to sets of vertices in opposite partitions of bipartite graphs, choosing $k$ so that $(\lambda/d)^k = (5\sqrt{2}/8)^k \leq \epsilon$, and setting $d_\epsilon = 2d^k$ so as to obtain Equation 1. For completeness, we state the Lemma below, with a short proof.

We remark that this is the first step in which we depart from [20], that instead proves a results similar to those in Subsections 3.2 and 3.3 relying directly on the eigengap. We believe using the Expander Mixing Lemma as an intermediate stepping point makes the subsequent proofs simpler and more intuitive (being based on the direct properties of graphs, rather than on the algebraic properties of their adjacency matrices), and in general our results more modular (as the spectral-gap route is not the only way to obtain graphs satisfying the simple pseudorandomness condition of Definition 1).

**Lemma 1.** *A $d_\epsilon-$regular bipartite multigraph $G_N = (L, R, E)$ with $|L| = |R| = N$ that has an irreducible biadjacency matrix with eigengap at least $\epsilon^{-1}$ satisfies property* **DISC**$(\epsilon d_\epsilon)$ *from Definition 1.*

*Proof.* Denote by $u_i$ and $v_i$, with $1 \leq i \leq N$, the generic node of $L$ and $R$, respectively. Denote by $\mathbf{A}$ the $N \times N$ biadjancency matrix of $G_N$, whose generic component $a_{i,j}$ equals the number of edges to $v_i$ from $u_j$; noting that since $G_N$ is $d_\epsilon-$regular the dominant eigenvalue of $\mathbf{A}$ is $d_\epsilon$, relative to the eigenvector $\mathbf{1}$ (the column vector whose components are all 1). For any subsets $U \subseteq L$ and $V \subseteq R$ denote by $\mathbf{X}_U$ and $\mathbf{X}_V$ the $N-$dimensional column vectors whose $i^{th}$ components equal 1 respectively if $u_i \in U$ and $v_i \in V$, and 0 otherwise (so $\mathbf{X}_L = \mathbf{X}_R = \mathbf{1}$); and given a generic vector $\mathbf{X}_W$, denote by $\mathbf{X}_W^=$ its projection on $\mathbf{1}$ and by $\mathbf{X}_W^\perp = \mathbf{X}_W - \mathbf{X}_W^=$ its orthogonal component. Note that $\mathbf{X}_W^= = (\frac{1}{\sqrt{N}}\mathbf{1}^T \cdot \mathbf{X}_W)\frac{1}{\sqrt{N}}\mathbf{1} = \frac{|W|}{N}\mathbf{1}$. Therefore:

$$(2) \quad e(U,V) = \mathbf{X}_V^T \mathbf{A} \mathbf{X}_U = \mathbf{X}_V^{=T} \mathbf{A} \mathbf{X}_U^= + \mathbf{X}_V^{\perp T} \mathbf{A} \mathbf{X}_U^\perp = \frac{|U|}{N}\mathbf{1}^T \mathbf{A} \frac{|V|}{N}\mathbf{1} + \mathbf{X}_V^{\perp T} \mathbf{A} \mathbf{X}_U^\perp$$

and thus:

$$(3) \quad \left|e(U,V) - \frac{|U||V|}{N^2}d_\epsilon N\right| = \mathbf{X}_V^{\perp T} \mathbf{A} \mathbf{X}_U^\perp \leq \|\mathbf{X}_V^\perp\|_2 \|\mathbf{A} \mathbf{X}_U^\perp\|_2 \leq \|\mathbf{X}_V^\perp\|_2 (\epsilon d_\epsilon \|\mathbf{X}_U^\perp\|_2)$$

as the eigengap of $\mathbf{A}$ is at least $\epsilon^{-1}$. And since for a generic $\mathbf{X}_W$ we have $(\|\mathbf{X}_W\|_2)^2 = \frac{|W|}{N}\mathbf{1}^T \cdot \frac{|W|}{N}\mathbf{1} + (\|\mathbf{X}_W^\perp\|_2)^2$, and thus $(\|\mathbf{X}_W^\perp\|_2)^2 = |W| - \frac{|W|^2}{N}$, then:

$$(4) \quad \left|e(U,V) - \frac{d_\epsilon}{N}|U||V|\right| \leq \epsilon d_\epsilon \sqrt{|U||V|(1-|U|/N)(1-|V|/N)} \leq \epsilon d_\epsilon \sqrt{|U||V|}.$$

□

3. Deterministic oblivious loose compaction in linear time

This section shows how to carry out, for any sufficiently large $\ell$, obliviously, deterministically and in linear time a slight variant (more amenable to composition) of $\ell$−loose compaction – namely:

**Definition 2.** *2-fold compaction at density $1/\ell$: given an array of $n$ words (of $b \geq \lg n$ bits) of which at most $m \leq n/\ell$ are marked, permute the words so that all the marked words end in the first half of the array.*

We assume hereafter for simplicity that $n$ is a power of 2. Note that if we can perform 2−fold compaction at density $1/\ell$ in time $T(n)$, we can easily perform $\ell$−loose compaction in time $2T(n)+n$: first we count the number $m$ of marked words in time $n$ with a simple linear scan, then perform 2−fold compaction iteratively on the first $n, n/2, \ldots, 2^{\lceil \lg_2(m\ell) \rceil}$ positions of the array, and finally "pretend" to perform 2−fold compaction (i.e. access the array locations without moving any words) on the first $2^{\lfloor \lg_2(m\ell) \rfloor}, \ldots, 1$ positions.

3.1. **Arrays and multigraphs.** The key tools for many of our tasks (both in this section, and in the next) are the families of pseudorandom graphs from Section 2; for each task, independently of the input size $n$, we choose a family satisfying the **DISC**($\epsilon d_\epsilon$) property for a sufficiently small $\epsilon$ – and then use graphs from that family whose size might depend on $n$ (but we stress $\epsilon$ and $d_\epsilon$ do not). Our strategy revolves around partitioning our arrays into a number of *blocks*, each holding an identical, $O(1)$ number of *atoms*. Atoms are in most cases individual words, but sometimes entire subarrays of words; in any case we treat atoms as indivisible objects that can only be copied or deleted atomically, in time proportional to the number of words in them. We work with pairs of $N$-block arrays (intuitively, one holds the inputs and the other the outputs), associating the $i^{th}$ block of the first array, and the $j^{th}$ block of the second, respectively to the $i^{th}$ vertex in one partition, and to the $j^{th}$ vertex in the other partition, of a multigraph with $N$ vertices per partition. We then speak interchangeably of blocks and vertices, and of arrays and partitions (see Figure 1). The key idea is to ensure that all our operations take place between atoms of adjacent vertices, considering each vertex in turn in a fixed order that does not depend on the contents of the elements. This guarantees obliviousness, and also linearity, since for each vertex there are at most $O(d_\epsilon) = O(1)$ pairs of atoms one from that vertex and one from a neighbour – so acting $O(1)$ times on all such pairs takes time proportional to scanning each array once.

3.2. **Matchings yield compaction.** As an application of the technique above, we show how to perform 2−fold compaction obliviously in time $O(n)$ if we can find a subset of edges with a certain property in one of the pseudorandom graphs from Section 2. Choose a family of $d_\epsilon$-regular multigraphs $G_{2^i}(L_{2^i}, R_{2^i}, E_{2^i})$ with $|L_{2^i}| = |R_{2^i}| = 2^i, i \in \mathbb{N}$, satisfying the **DISC**($\epsilon d_\epsilon$) property for $\epsilon \leq 1/64$. Let $B$ be the largest power of 2 no larger than $d_\epsilon/2$. Partition the main array into $N = n/B$ blocks of $B$ word-atoms, and similarly partition an auxiliary array of identical size filled with dummy words; then associate the blocks to the vertices of $G_N$. Call a vertex marked if it contains more than $B/4$ marked words, and let $L' \subseteq L_N$ be the set of such vertices. Our goal is to find a $(B, B/4)$-*matching* for $L'$:

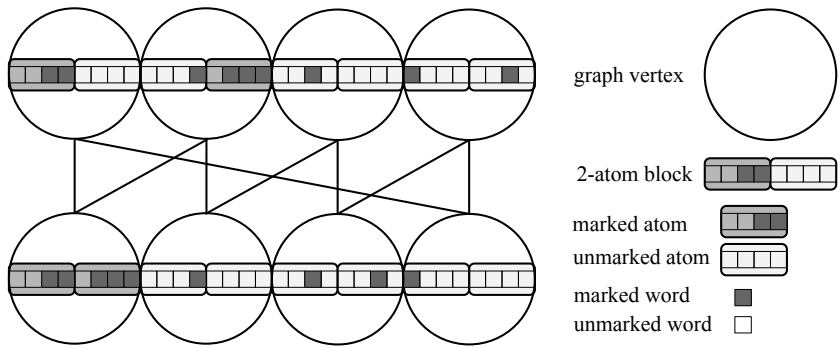

FIGURE 1. The two partitions of a bipartite (multi)graph are associated to two permutations of the same array. Each vertex is associated to a block of 2 atoms, each atom being a subarray of 4 words. Note that each atom and its image are always in adjacent vertices.

**Definition 3.** *Given a bipartite multigraph $G = (L, R, E)$ and a subset $L' \in L$, an $(a, b)$ matching for $L'$ is a set of edges $E' \subseteq E$ such that each vertex of $L'$ has at least $a$ incident edges in $E'$, and each vertex of $R$ has at most $b$.*

We shall hereafter speak of atoms, instead of words, since the argument works with multi-word atoms too (something we exploit later on). Note that any array with at most $\gamma/4$ marked atoms contains at most $\gamma$ marked blocks. Then, if we can find, obliviously and in time $T(N)$, a $(B, B/4)$-matching for any arbitrary set $L'$ of at most $\gamma N$ marked vertices, we can obliviously perform 2−fold compaction at density $\gamma/4$ in time $T(N) + O(N)$ in three simple steps. First, we swap the marked atoms of each marked vertex in $L'$ with dummy atoms of matched neighbours in $R_N$ – each of which will receive at most $B/4$ marked atoms *from all its own neighbours*. As a second step, we swap all marked atoms in the $(i + N/2)^{th}$ block of $L_N$ and in the $i^{th}$ and $(i + N/2)^{th}$ blocks of $R_N$ (which are at most $3B/4$) with unmarked atoms in the $i^{th}$ block of $L_N$ (which are at least $B - B/4 = 3B/4$), leaving only unmarked and dummy atoms in $R_N$ and in the $N/2$ higher-index vertices of $L_N$. Finally, noting that in the second step we did not move any dummy atoms, we have each vertex $L_N$ trade back any dummy atoms to its neighbours of $R_N$ that provided them in the first step – leaving all dummy atoms in $R_N$, and all marked atoms in the lower-index vertices of $L_N$. We can perform all the swaps in the first step obliviously in time $O(B^2 d_\epsilon N) = O(N)$ simply considering (each pair of atoms in) each pair of neighbours in $G_N$ in turn; this also means we can perform all the swaps in the third step obliviously in time $O(N)$, and obviously we can perform all those in the second step obliviously in time $O(B^2 N) = O(N)$.

3.3. **Non-oblivious matchings.** To compute the matching non-obliviously, but in time $O(N)$, we can run sequentially the simple distributed protocol of [20] that has each marked vertex determine the matching edges incident on it. A marked vertex that has not yet done so is *unsatisfied*; over a number of rounds, unsatisfied vertices exchange messages with their neighbours and eventually all become satisfied – at which point the protocol terminates. More precisely every round, for every edge $(u, v)$ incident on an unsatisfied vertex $u$, a request is sent from $u$ to $v$; then, every vertex in $R_{N/B}$ that has received that round more than $B/4$ requests replies negatively *to each*, and every vertex that has received at most $B/4$ replies positively; finally, every unsatisfied vertex $u$ that sees at least $B$ positive replies takes the corresponding edges as a matching and becomes satisfied.

It is immediate to see that every round each edge exchanges at most $O(1)$ bits, and each vertex performs $O(1)$ elementary operations – in fact, only unsatisfied vertices and their neighbours must perform any operation at all. It is also easy to see that no vertex in $R_{N/B}$ replies positively *over all* rounds to more than $B/4$ requests: the first round in which it replies positively, it has received no more than $B/4$ requests, and those requests are a superset of those it will receive in subsequent rounds (an edge $(u,v)$ stops forwarding requests from $u$ only once $u$ is satisfied, and at that point it effectively falls silent). It is only slightly more difficult to prove:

**Lemma 2.** *If $\gamma \leq \frac{1}{32}$, unsatisfied vertices decrease by a factor at least $2$ each round.*

*Proof.* Let $U$ be the set of unsatisfied vertices at beginning of any given round, and let $W$ be a generic set of neighbours such that $e(U,W) > |W|B/4$. From Equation 1, setting $V = W$, we obtain $|W|B/4 < e(U,W) \leq d_\epsilon|U||W|/(N/B) + \epsilon d_\epsilon\sqrt{|U||W|}$; dividing by $|W|d_\epsilon$ and rearranging this becomes $\epsilon\sqrt{|U|/|W|} \geq B/(4d_\epsilon) - |U|/|L_{N/B}|$. Remember that we chose $\epsilon \leq 64$, and $B$ as the largest power of $2$ no larger than $d_\epsilon/2$, so $B/d_\epsilon > 1/4$; then if $|U|/|L_{N/B}| \leq 1/32$ we immediately have $\sqrt{|U|/|W|} \geq 64/16 - 64/32$, i.e. $|U|/|W| \geq 4$. Then, at most one quarter of all the edges in $U$ will belong to $W$ and yield a negative reply; and then at least half the nodes in $U$ see at least half their requests (i.e. at least $d_\epsilon/2 \geq B$) answered positively, and become satisfied. □

By Lemma 2, obtaining a matching takes $O(\lg N)$ rounds, and a total of $O(N)$ timesteps since the number of unsatisfied vertices that must be considered every round decreases geometrically. However, the resulting scheme is *not* oblivious, since the sets of vertices visited after the first round depend on the input.

**Remark:** Alternatively, the matching can be obtained obliviously by examining every round *all* the vertices, and not only the unsatisfied ones, but this increases the cost to $\Theta(N \lg N)$ (nonetheless, this will prove useful in Subsection 3.6).

It might then seem we have spent significant effort for little progress. In fact, we have decoupled the actual movement of the data we want to compact (a task that can be carried out obliviously in linear time), from the computation of how to compact it. We exploit this advantage, and in particular the fact that the number of bits per atom used to compute the matching can be significantly lower than the number of bits *in* each atom, in the next three subsections.

3.4. **Oblivious matchings for** $n = O(b/\lg b)$. If we work with a sufficiently small graph, we can compress all the information necessary to determine the matching into $O(1)$ words, and achieve obliviousness without asymptotically increasing cost. In particular, we can represent each $N$-vertex partition of a $d_\epsilon$-regular balanced bipartite graph as a sequence of $O(Nd_\epsilon)$ pointers of $\lg(N)$ bits each (one for each edge, to one of the $N$ vertices in the other partition), for a total of $O(N \lg N)$ bits. With no asymptotic increase in the number of bits, we can also label each vertex with a number between $0$ and $d_\epsilon$ (corresponding to the number of requests/of positive answers received) and each edge-field with flags indicating the status of the request, its answer, and whether the edge is now part of a matching – and we can also maintain a progressively shrinking list of pointers to unsatisfied vertices. It is immediate to verify that computing the matching on the compressed representation (and subsequently reading it to perform the actual word transfer) requires $O(N)$ elementary arithmetic and bit-shift operations. We can then perform 2−fold compaction at density $\gamma/4$ obliviously and in time $O(n)$ on any array of size $n = O(b/\lg b)$, where $b$ is the number of bits per word and thus $2^b$ is the size of the addressable memory.

3.5. **Oblivious compaction for** $n = O((b/\lg b)^2)$**.** If $n = \omega(b/\lg b)$ (otherwise we can simply apply the scheme above) but also $n = O((b/\lg b)^2)$, let $p, q$ be two powers of 2 such that $p \leq q = \Theta(b/\lg b)$. Consider an array of $n = pq$ words, partitioned into $p$ subarrays of $q$ words each. If each subarray contained at most $q(\gamma/4)$ marked words, we could compact each independently, producing an array of $2p$ "half-subarrays" with all marked words in the odd ones (the first, third etc.). This would yield 2−fold compaction in time $O(n)$ of the entire array since in general, given an array of $a \cdot b$ elements (words, blocks, etc.) of which only those with index $i \equiv 1 \mod a$ are marked, we can easily permute it obliviously so that only the first $b$ elements are marked – simply by treating the array as a $a \times b$ 2-dimensional array stored column-wise, that we then read row-wise, finding only the first row marked.

Unfortunately, some subarrays can contain significantly more marked words than the global average; on the other hand, such subarrays cannot be too many. More precisely, assume the fraction of marked words in the array is at most $(\gamma/8)^2$, and let a subarray be *marked* if it contains a fraction of marked words greater than $\gamma/8$; thus the fraction of marked subarrays cannot exceed $\gamma/8$. We can then apply the 2−fold compaction scheme of Subsection 3.4 with each subarray as an atom, so our array has size *measured in atoms* equal to $p = O(b/\lg b)$. In fact, since the fraction of marked atoms is at most $\gamma/8$ rather than $\gamma/4$, we can perform compaction *twice*, moving all the marked atoms to the first $1/4$ of the array – all in time $O(n)$. The remaining $3/4$ of the array contain only subarrays each with a fraction of marked words no higher than $\gamma/8$; these subarrays can the be compacted individually with words as atoms – and again twice, so that in time $O(n)$ we can then move all their marked words into the first $1/4$ of the second $3/4$ of the array (see above). This ensures all marked words end in the first half of the array, yielding oblivious 2−fold compaction at density $(\gamma/8)^2 = 2^{-16}$ in time $O(n)$ as long as $n = O((b/\lg b)^2)$.

3.6. **Oblivious compaction for arbitrary** $n$**.** The trick of "zooming out", treating entire subarrays of $\Theta(b/\lg b)$ words as single atoms, allows us to square the maximum compactable array size – but at the cost of squaring the maximum tolerable mark density, so we cannot zoom out too many times. However, once we can compact subarrays of size $\Omega((b/\lg b)^2) = \Omega(b)$ as in the previous subsection, we can enact a variant of the trick *once* to compact arrays of arbitrary size as long as the fraction of marked words is sufficiently low.

More precisely, consider an array of arbitrary size $n$, with a fraction of marked words no higher than $(\gamma/8)(\gamma/8)^2/2 = 2^{-25}$. If $n = \omega((b/\lg b)^2)$ (otherwise we can simply apply the compaction scheme of the previous subsection), let $p$ be a power of 2 such that $p = \Theta((b/\lg b)^2)$, and partition the main array into $O(n/p)$ subarrays of size $p$. Let any such subarray be *marked* if it contains a fraction of marked words greater than $(\gamma/8)^2/2$; then the fraction of marked subarrays cannot exceed $\gamma/8$. Treating each subarray as an atom, the associated multigraph has $N = \Theta(n/p)$ vertices; we can then compute a matching on it obliviously in time $O(N \lg N)$, for arbitrarily large $N$ by going through *all* vertices at each of the $\lg N$ stages of the computation, rather than considering only unsatisfied ones (see Subsection 3.2, in particular the remark below Lemma 2). And since $p = \Omega(b) = \Omega(\lg n)$ and thus $N \lg N = O((n/p)(\lg n)) = O(n)$, and the fraction of marked atoms is at most $\gamma/8$, we can compact all marked atoms, twice, into the first $1/4$ of the array in time $O(n)$. Next, we simply apply the compaction scheme of Subsection 3.5 to each of the remaining subarrays, twice, and move the marked words into the first $1/4$ of the second $3/4$ of the array – all in time $O(n)$. We have then proved:

**Lemma 3.** *Deterministic oblivious* 2-*fold compaction at density* $1/\ell$ *(and thus* $\ell-$*loose compaction) of any array of size* $n$ *for* $\ell = 2^{25}$ *requires time* $O(n)$.

**Remark:** the 2−fold compaction scheme of this section copies a permutation of the contents of an input array $A[\ ]$ into an output array $B[\ ]$, computing independently of the contents of $A[\ ]$ a sequence of index pairs $(a_i, b_i)$, and then acting in turn on every pair of words $A[a_i], B[b_i]$ either to copy $A[a_i]$ into $B[b_i]$, or to leave both words unchanged (a choice that *does* depend on the contents of $A[\ ]$). Denote by $\pi(h)$ the index, in $B[\ ]$, of the word initially in $A[h]$. Then, if we use the compaction scheme to copy a permutation of $A[\ ]$ into $B[\ ]$, and then process some words in $B[\ ]$ obtaining an array $B'[\ ]$, we can import the changes effected on $B[\pi(h)]$ back into $A[h]$ simply by going through the same sequence of index pairs, and copying $B'[b_i]$ into $A[a_i]$ if and only if $A[a_i]$ was copied into $B[b_i]$. This *failure sweeping* [10] process is a key tool for the main construction in the next section.

## 4. Deterministic oblivious distribution in linear time

Consider an array of $n$ words, with $m$ marked words and $m$ marked positions, assuming without loss of generality that $n$ is a power of 2 (otherwise, we can imagine extending the array with dummy unmarked words and dummy unmarked positions). Call marked words in unmarked positions, and unmarked words in marked positions, *mismatched*; imagine the first are coloured red, and the second blue, noting that they are equal in number. To permute the array so that all marked words end in marked positions, all we have to do is swap every blue word with a (distinct) red word. In fact, a routine that swaps all the mismatched words *save at most* $n/\ell$ in time $O(n)$ (where $\ell$ is the constant from Lemma 3) automatically yields an algorithm to swap *all* the mismatched words, none excluded, in time $T(n) = O(n)$: the algorithm first uses the routine to swap "almost all" the mismatched words, in time $O(n)$; then it invokes the loose compaction scheme of Section 3 to obliviously move the remaining mismatched words (let us call them the *survivors*) into an array of size $n/2$, in time $O(n)$; then it recursively calls itself to swap *all* the survivors, now in the smaller array, in time $T(n/2)$; and finally invokes compaction scheme in reverse (see the remark at the end of the previous section) to obliviously move the survivors, that have all been swapped through the recursion, back into the main array. The algorithm is obviously deterministic and oblivious, and sports a running time $T(n) = T(n/2) + O(n) = O(n)$ (with the base case of the recursion acting on an array of size $O(1)$, which obviously requires time $O(1)$).

All we need is then a routine that, in any array of size $n$ with $m$ red words and $m$ blue words, swaps every red word with a blue word, save possibly for up to $s = n/(2\ell)$ red and $s$ blue survivors. Consider a family of $d_\epsilon$-regular multigraphs $G_{2^i}(L_{2^i}, R_{2^i}, E_{2^i})$ with $|L_{2^i}| = |R_{2^i}| = 2^i$, $i \in \mathbb{N}$, satisfying the **DISC**$(\epsilon d_\epsilon)$ property for $\epsilon = \frac{1}{2\sqrt{\ell}}$; and associate to each position of a generic array of size $n$ a distinct vertex in $L_n$, allowing us to speak interchangeably of array positions and of vertices in $L_n$. The routine consists simply in visiting, for each vertex in $R_n$, all its $\binom{d_\epsilon}{2}$ pairs of neighbours, swapping the words in any given pair if and only if one is red, one is blue, and neither has yet been swapped. The order in which the pairs are visited can be arbitrary (as long as it is independent of the contents of the array).

It is immediate that the routine takes $n\binom{d_\epsilon}{2} = O(n)$ oblivious steps. All we have to prove is that it leaves at most $s$ blue and $s$ red survivors. Obviously, a red and a blue survivor cannot reside in two vertices sharing a common neighbour. But we can easily show that every set of more than $s$ vertices has more than $n/2$ neighbours, so no two such sets can be disjoint. To this end, consider Equation 1, letting $U$ be a generic subset of $L_n$, and $V$ the set of its neighbours (so $e(U,V) = d_\epsilon|U|$). Then $|d_\epsilon|U| - d_\epsilon|U||V|/n| \le \frac{1}{2\sqrt{\ell}}\sqrt{|U||V|}$, i.e. $|1 - |V|/n| \le \sqrt{|V|/(4\ell|U|)}$; and the last inequality, if we had both $|V| \le n/2$ and $|U| > n/(2\ell)$, would yield $|1 - 1/2| < \sqrt{(n/2)/(4\ell n/2\ell)}$ i.e. $1/2 < 1/2$. We have finally proved:

**Theorem 1.** *Given a generic array of $n$ elements, with $m$ positions and $m$ elements marked, one can permute the contents of the array so that every marked element ends in a marked position, with a deterministic sequence of $O(n)$ memory accesses independent of the contents of the array.*

## 5. Conclusions

We have shown how to perform deterministic oblivious distribution in linear time in the standard word-RAM model – asymptotically improving to an optimal $O(n)$ the running time for deterministic tight compaction, randomized tight compaction, and randomized $O(1)$−loose compaction of $n$−element arrays.

While our running time is indeed $O(n)$, we freely admit that the constants hidden in the asymptotic notation are *very* large: simulating a butterfly network with $(n\lg n)/2$ oblivious compare-and-swaps appears more efficient for all realistic $n$. There are two main reasons behind the large constants. The first is that we have sacrificed constant factors rather liberally in our construction (mostly to keep quantities aligned to powers of 2, and to avoid analysing too many cases when packing data) so as to keep the overall picture as simple as possible; mitigating these losses is not difficult if one is willing to accept a dirtier solution. The second and major reason is the use of families of pseudorandom graphs that are, fundamentally, expanders. In fact, the requirement that these families be sufficiently dense and explicitly constructible in linear time (at least without too many complications) pushed us towards older, simpler constructions with weaker expansion properties. "Better" expanders such as Ramanujan graphs [17] would significantly reduce constants, though the Alon-Boppana bound [19, 6] does present a hard limit to how low we can keep them for a given $\epsilon$ in Equation 1. Note that we do not need Equation 1 to hold for *every* subset of vertices, but only for some; this may be a key to reducing constants further, perhaps trading back determinism for practical performance.

A second direction to explore, in terms of practicality of our scheme, is its performance under memory models more realistic than the RAM that incorporate e.g. parallelism, non-uniform access cost, block transfer, and pipelining. It is not too difficult to show that our scheme parallelizes well, and in particular runs in time $O(n/p)$ on any PRAM with $p = O(n/\lg n)$ processors – which automatically translates into good pipelined performance. Caching and, more in general, non-uniform access cost is a non-isse in terms of asymptotics since we access each datum $O(1)$ times; on the other hand it may well have an impact on the constants involved, and help with block transfer. Dealing with block transfer appears, in fact, the main challenge: roughly speaking, dense sets of words from a source array should map into sparse sets in the destination array to guarantee good compaction performance, whereas they should map into dense sets to guarantee good performance under block transfer. Sufficient non-uniformity (informally, a sufficiently large cache) and/or pipelining can mitigate this conflict [4, 10]; but exactly to what extent and under which parameter choices are issues that merit further research.

## References

[1] N. Alon. Eigenvalues and expanders. *Combinatorica*, 6(2):83–96, 1986.
[2] K. E. Batcher. Sorting networks and their applications. In *Proc. of AFIPS*, pages 307–314, 1968.
[3] M. Blanton, A. Steele, and M. Aliasgari. Data-oblivious graph algorithms for secure computation and outsourcing. In *Proc. of ASIA CCS*, pages 207–218, 2013.
[4] T. H. Chan, Y. Guo, W. Lin, and E. Shi. Cache-oblivious and data-oblivious sorting and applications. In *Proc. of SODA*, pages 2201–2220, 2018.
[5] T. H. Chan, K. Nayak, and E. Shi. Perfectly secure oblivious parallel RAM. *IACR Cryptology ePrint Archive*, 2018:364, 2018.
[6] F. Chung. A generalized Alon-Boppana bound and weak Ramanujan graphs. *Electronic Journal of Combinatorics*, 23(3):207 – 210, 2016.
[7] F. Chung and R. Graham. Sparse quasi-random graphs. *Combinatorica*, 22(2):217–244, 2002.
[8] F. Chung and R. Graham. Quasi-random graphs with given degree sequences. *Random Struct. Algorithms*, 32(1):1–19, 2008.
[9] O. Goldreich and R. Ostrovsky. Software protection and simulation on oblivious RAMs. *J. ACM*, 43(3):431–473, 1996.
[10] M. T. Goodrich. Data-oblivious external-memory algorithms for the compaction, selection, and sorting of outsourced data. In *Proc. of SPAA*, pages 379–388, 2011.
[11] M. T. Goodrich and M. Mitzenmacher. Privacy-preserving access of outsourced data via oblivious RAM simulation. In *Proc. of ICALP*, pages 576–587, 2011.
[12] T. Hagerup. Sorting and searching on the word RAM. In *Proc. of STACS*, pages 366–398, 1998.
[13] S. Jimbo and A. Maruoka. Expanders obtained from affine transformations. *Combinatorica*, 7(4):343–355, 1987.
[14] Y. Kohayakawa, V. Roedl, M. Schacht, P. Sissokho, and J. Skokan. Turan's theorem for pseudo-random graphs. *Journal of Combinatorial Theory, Series A*, 114(4):631 – 657, 2007.
[15] E. Kushilevitz, S. Lu, and R. Ostrovsky. On the (in)security of hash-based oblivious RAM and a new balancing scheme. In *Proc. of SODA*, pages 143–156, 2012.
[16] T. Leighton, Y. Ma, and T. Suel. On probabilistic networks for selection, merging, and sorting. *Theory of Computing Systems*, 30(6):559–582, 1997.
[17] A. Lubotzky, R. Phillips, and P. Sarnak. Ramanujan graphs. *Combinatorica*, 8(3):261–277, 1988.
[18] K. Nayak, X. S. Wang, S. Ioannidis, U. Weinsberg, N. Taft, and E. Shi. GraphSC: Parallel secure computation made easy. In *Proc. of IEEE Security & Privacy*, pages 377–394, 2015.
[19] A. Nilli. On the second eigenvalue of a graph. *Discrete Mathematics*, 91(2):207 – 210, 1991.
[20] N. Pippenger. Self -routing superconcentrators. *Journal of Computer and System Sciences*, 52(1):53 – 60, 1996.
[21] L. G. Valiant. A scheme for fast parallel communication. *SIAM J. Comput.*, 11(2):350–361, 1982.